# Novel features of Josephson flux-flow in Bi-2212: contribution of in-plane dissipation, coherent response to mm-wave radiation, size effect

Yu. I. Latyshev[1,2], A. E. Koshelev[3], V. N. Pavlenko[*,1,2], M. Gaifullin[4], T. Yamashita[1], and Y. Matsuda[4]

[1]RIEC, Tohoku Univ., Katahira, Aoba-ku, Sendai, 980-8577, Japan
[2]Institute of Radio-Engineering and Electronics Russian Acad. of Sci., Mokhovaya 11-7, Moscow 101999, Russia
[3]Materials Science Division, Argonne National Laboratory, Argonne, Illinois 60439, USA
[4]Institute for Solid State Physics, Univ. of Tokyo, Kashiwanoha 5-1-5, Kashiwa, Chiba 277-8581, Japan

**Abstract.** We studied Josephson flux-flow (JFF) in Bi-2212 stacks fabricated from single crystal whiskers by focused ion beam technique. For long junctions with the in-plane sizes 30 x 2 $\mu m^2$, we found considerable contribution of the in-plane dissipation to the JFF resistivity, $r_{Jff}$, at low temperatures. According to recent theory [A. Koshelev, Phys. Rev. **B62**, R3616 (2000)] that results in quadratic type dependence of $r_{Jff}(B)$ with the following saturation. The *I-V* characteristics in JFF regime also can be described consistently by that theory. In JFF regime we found Shapiro-step response to the external mm-wave radiation. The step position is proportional to the frequency of applied microwaves and corresponds to the Josephson emission from all the 60 intrinsic junctions of the stack being synchronized. That implies the coherence of the JFF over the whole thickness of the stack and demonstrates possibility of synchronization of intrinsic junctions by magnetic field. We also found a threshold character of an appearance of the JFF branch on the *I-V* characteristic with the increase of magnetic field, the threshold field $B_t$ being scaled with the junction size perpendicular to the field L (L = 30–1.4 $\mu$m), as $B_t » F_0 / L\, s$, where $F_0$ is flux quantum, *s* is the interlayer spacing. On the *I-V* characteristics of small stacks in the JFF regime we found Fiske-step features associated with resonance of Josephson radiation with the main resonance cavity mode in transmission line formed by stack.

**PACS numbers:** 74.72.Hs, 74.60.Ge.

**Keywords:** Josephson flux-flow, Josephson emission, Fiske steps.

---

[*] Corresponding author. Tel.: +7(095)203-4976; fax: +7(095)203-8414; e-mail: vit@iname.com

**Introduction**

   Studies of interlayer tunnelling in layered high-$T_c$ materials associated with intrinsic Josephson effects [1,2] lately became one of the most interesting issues developed in high-$T_c$ superconductivity. Existence of the intrinsic Josephson effects and the related Josephson plasma oscillations are now quite well documented [3]. Of a particular interest in this field is dynamics of Josephson vortex lattice (JVL) in long stacked junctions. JVL is formed in magnetic field parallel to the layers [4] and can be driven by dc current across the layers. Experimentally that was indicated as a resistive upturned flux-flow branch in the *I-V* characteristics with maximum voltage being proportional to magnetic field [5-7]. Maximum voltage corresponds to a condition, when the JVL velocity approaches the Swihart velocity [8], the velocity of electromagnetic wave propagation in the structure.

   As it was pointed out recently [9] for highly anisotropic materials, as Bi-2212, the flux-flow resistivity at low temperatures can contain considerable contribution of the in-plane quasiparticle dissipation and thus can be used for extraction the information about both quasiparticle conductivity components in superconducting state. This method still is not well elaborated. The coherent motion of JVL should induce coherent Josephson emission as it occurs in conventional long Josephson junctions [10]. The emission of that type can provide useful information about coherence of JVL motion. Therefore experiments on detection of Josephson emission in sliding JVL are very important. The interesting point in view of possible applications is also the short length limit for an observation of flux-flow regime. The present paper has been addressed to clarify the most of questions listed above. We have undertaken the detailed studies of the *I-V* characteristics and flux-flow resistivity in the sliding JVL regime. The data have been shown to be well consistent with recent theories of sliding JVL. We found Shapiro step-type response of sliding JVL state to the external microwaves and Fiske step features both pointing out to the existence of coherent Josephson emission in sliding JVL regime. The emission corresponds to the vertical synchronization of the major part or in some cases of the all number of elementary junctions in the stack. We also have studied flux-flow behaviour with a decrease of the stack length down to ~1 micron.

**2. Experimental**

   The stacked structures have been fabricated by double-sided processing of high quality Bi-2212 whiskers [11] by focused ion beam (FIB). The stages of fabrications were similar to ones described in [12]. Fig. 1a shows schematically the geometry and orientation of the structure with respect to the crystallographic axes. The structure sizes were $L_a$ =1.5-30 µm, $L_b$ = 1-2 µm, $L_c$ = 0.05-0.15 µm. The sizes and other parameters of the structures used in our studies are listed in Table 1. The oxygen doping level of the stacks estimated from $r_c(T)$ measurements above $T_c$ [13] was nearly optimum, δ ≈ 0.25. The critical current density $J_c$ at 4.2 K in the absence of magnetic field was 1- 2 kA/cm$^2$ for the best samples.

   Measurements of the *I-V* characteristics of Bi-2212 stacks have been carried out in commercial cryostat of Quantum Design PPMS facility. The magnetic field has been oriented parallel to the *b*-axis within accuracy 0.1°. Field has been changed in steps of 0.1 T. In each fixed value of the field the back and forth *I-V* characteristics have been measured using fast oscilloscope. For the microwave measurements we used more advanced set up, the cryostat with slit pair superconducting magnet in the shielded room. The accuracy of magnetic field orientation B // b was 0.01°. External microwaves with a maximum incident power of 35 mW at the frequencies ranging from 45 to 142 GHz were applied from the backward-wave-oscillators. The samples were mounted on the substrate that was placed at the center of the rectangular waveguides and was capacitively coupled with the flange of a waveguide. Thus the electrical component has been kept parallel to the *c*-axis.

## 3. Flux-flow resistivity and the *I-V* characteristics

Fig. 1b shows a set of the *I-V* characteristics of a long stack #4 at T = 20 K with subsequent increase of magnetic field B oriented parallel to the b-axis. The flux-flow branch and its evolution with magnetic field are clearly seen in the picture. The flux-flow branch appears just above $I_c$ on the *I-V* characteristic. That is characterized by upturn and the maximum voltage, $V_{Jff}$, above which the switch to the multibranched state occurs. As known [5-7] $V_{Jff}$ linearly grows with field. We found the coefficient $dV_{Jff}/dB$ to be 0.42 mV/T per elementary junction. Above approximately 1.7 T the slope of $V_{Jff}(B)$ gradually increases to 0.86 mV/T (see also [6]). We define Josephson flux-flow resistance, $R_{Jff}$, as initial linear part of the flux-flow branch. That can be found from linear extrapolation of the *I-V* to $V \to 0$. This extrapolation gives more reliable values for high fields above 0.3-0.5 T, when $I_c(B)$ becomes small. In the following two sections we consider the results of calculations of flux-flow resistivity dependence on magnetic field and the results of numerical calculations of flux-flow branch on the *I-V* characteristics, both based on the recent advanced approach [9,14,15] that takes into account the in-plane dissipation channel. Note that in earlier calculations [16-18] this contribution has been ignored. Comparison with experiment presented below proved the importance of the in-plane dissipation channel in Josephson flux-flow dynamics in Bi-2212.

*3.1. Flux-flow resistivity*

The linear flux-flow resistivity of the Josephson vortex lattice, $r_{Jff}$, is determined by the static lattice structure and linear quasiparticle dissipation. At high fields, $B > \Phi_0/\pi\gamma s^2$ ($\gamma$ is the anisotropy of London penetration depths, $\gamma = \lambda_c/\lambda_{ab}$), Josephson vortices homegeneously fill all the layers and the static lattice structure is characterized by oscillating patterns of both *c*-axis and in-plane supercurrents. At small velocities this pattern slowly drifts along the direction of layers, preserving its static structure. This motion produces oscillating *c*-axis $(\tilde{E}_z)$ and in-plane $(\tilde{E}_x)$ electric fields leading to extra dissipation, in addition to usual dissipation due to the dc electric field $E_z$. Total dissipation per unit volume is given by

$$\sigma_{Jff} E_z^2 = \sigma_c E_z^2 + \sigma_c \langle \tilde{E}_z^2 \rangle + \sigma_{ab} \langle \tilde{E}_x^2 \rangle \tag{1}$$

where $\sigma_{Jff} = 1/r_{Jff}$ is the flux-flow conductivity, $\langle .... \rangle$ means time and space average, $\sigma_c = 1/r_c$ and $\sigma_{ab} = 1/r_{ab}$ are the c-axis and in-plane quasiparticle conductivities. An expansion with respect to Josephson current at high fields allows to derive a simple analytical formula for the flux-flow resistivity [9,14,15]:

$$\boldsymbol{r}_{Jff}(B) = \frac{B^2}{B^2 + B_s^2} \boldsymbol{r}_c, \quad B_s = \sqrt{\frac{\boldsymbol{s}_{ab}}{\boldsymbol{s}_c}} \frac{\Phi_0}{\sqrt{2}\boldsymbol{pg}^2 s^2} \qquad (2)$$

A relative importance of the in-plane and c-axis dissipation channels is determined by the dimensionless ratio $\boldsymbol{s}_{ab}/\boldsymbol{s}_c \boldsymbol{g}^2$. In high-$T_c$ superconductors at low temperatures, T < 60 K, this ratio is large $\boldsymbol{s}_{ab}/\boldsymbol{s}_c \boldsymbol{g}^2 = 20$–50, i.e., the in-plane channel dominates.

Eq. (2) describes very well the experimental field dependence of the flux-flow resistivity (Fig. 2). Fit of the experimental dependence $\boldsymbol{r}_{Jff}(T)$ by Eq. 2 at 4.2 K gives $\boldsymbol{r}_c \approx 480$ Ω cm and $B_s = 3.3$ T. This allows to extract the combination $\sqrt{\boldsymbol{s}_{ab}/\boldsymbol{s}_c \boldsymbol{g}^4} \approx 0.017$. Using for critical current density at low temperatures the value 1.7 kA/cm$^2$ we can estimate $\boldsymbol{g}$ as $\boldsymbol{g}$=500. That gives an estimate for $\boldsymbol{s}_{ab}$ (4.2 K) ≈ 40 (mΩ cm)$^{-1}$ which is quite close to the value $\boldsymbol{s}_{ab} \approx 60$ (mΩ cm)$^{-1}$ found at low temperatures from the microwave experiments [19].

Fig. 2 shows that experimental dependence $R_{Jff}(B)$ is well described by the theory in wide temperature region 4.2 –60 K. From the fit of experimental curves to Eq. (2) we can extract the temperature dependence of $B_s$ and $\boldsymbol{s}_c$ (see insert to Fig.2). As shown in insert, a dependence $\boldsymbol{s}_c(T)$ found in this way well agrees with dependence for $\boldsymbol{s}_c(T)$ extracted from the independent experiment on small mesas in zero magnetic field [20]. To extract the temperature dependence of $\boldsymbol{s}_{ab}$ we need additional knowledge of the temperature dependence of $\boldsymbol{g}$. In principle that can be extracted from the temperature dependence of linear part of $R_{Jff}(B)$ dependences, but unfortunately the low field data of $R_{Jff}(B)$ are not so reliable.

To demonstrate the importance of in-plane contribution in flux-flow resistivity we show in Fig.2 two theoretical dependences at 4.2 K, one is fitted to the experiment with finite $\boldsymbol{s}_{ab}$ and another one is calculated with the same fitting parameters, but with zero in-plane dissipation contribution ($\boldsymbol{s}_{ab} = 0$). One can see huge inconsistency with experiment in the latter case.

*3.2. Flux-flow branch*

At high fields in the resistive state the interlayer phase differences grow approximately linearly in space and time

$$\boldsymbol{q}_n(t,x) \approx \boldsymbol{w}_E t + k_H x + \boldsymbol{f}_n, \qquad (3)$$

where $\boldsymbol{w}_E$ is the Josephson frequency and $k_H$ is magnetic wave vector. In the following we will use reduced parameters: $\boldsymbol{w} \to \boldsymbol{w}_E/\boldsymbol{w}_p$, $k_H \to k_H \boldsymbol{g} s$ (see Table 2). The most important degrees of freedom in this state are the phase shifts $\boldsymbol{f}_n$, which describe the structure of the moving Josephson vortex lattice. In particular, for the static triangular lattice $\boldsymbol{f}_n = \boldsymbol{p} n$. Lattice structure experiences a nontrivial evolution with increase of velocity. The equations for $\boldsymbol{f}_n$ can be derived from the coupled sine-Gordon equations for $\boldsymbol{q}_n(t,x)$ by expansion with respect to the Josephson current and averaging out fast degrees of freedom. In the case of steady state for a stack consisting of N junctions, this gives the following set of equations:

$$\frac{1}{2}\sum_{m=1}^{N} \text{Im}\left[\mathsf{g}(n,m)\exp\left(i(\boldsymbol{f}_m - \boldsymbol{f}_n)\right)\right] = i_J, \qquad (4)$$

where $i_J \equiv i_J(k_H, w_E) = \langle \sin q_n(t,x) \rangle$ is the reduced Josephson current, which has to be obtained as solution of these equations. The complex function $g(n,m)$ describes phase oscillations in the *m*-th layer excited by oscillating Josephson current in the *n*-th layer. For a finite system it consists of the bulk term $G(n-m)$ and top and bottom reflections (multiple reflections can be neglected):

$$g(n,m) = G(n-m) + BG(n+m) + BG(2N+2-n-m),$$

where

$$G(n) = \int \frac{dq}{2p} \exp(iqn) \left( w^2 - in_c w - \frac{k^2(1+in_{ab}w)}{2(1-\cos q) + (1+in_{ab}w)/l^2} \right)^{-1}, \quad (5)$$

$w = w_E$ and $k = k_H$ are the frequency and the in-plane wave vector of the travelling electromagnetic wave generated by moving lattice. Dissipation parameters, $n_c$ and $n_{ab}$, and reduced penetration depth *l* are defined in Table 2. $B = B(k,w)$ is the amplitude of reflected electromagnetic wave. For the practical case of the boundary between the static and moving Josephson lattices a detailed calculation of $B(k,w)$ is presented in Ref. [15]. In general, quasiparticle conductivities in definitions of $n_c$ and $n_{ab}$ are the complex conductivities at the Josephson frequency. The frequency dependence is especially important for the in-plane conductivity. Recent terahertz spectroscopy measurements of $s_{ab}(w)$ in BSCCO by Corson *et al.* [21] showed at low temperatures it has Drude frequency dependence with typical relaxation rate $1/t \approx 1$ THz. Maximum Josephson frequency at the termination point of the flux-flow branch exceeds this value at fields $\geq 2$ T. Therefore the frequency dependence has to be taken into account. We use the Drude-like frequency dependence of $n_{ab} \equiv n_{ab}(w)$:

$$n_{ab}(w) = \frac{n_{ab0}}{1+iwt} \quad (6)$$

where *t* is the quasiparticle relaxation time. Solution of Eq. (4) allows to obtain current-voltage characteristic as

$$j(E_z) = s_c E_z + j_J i_J(k_H, w_E),$$

where $k_H$ and $w_E$ has to be expressed via magnetic and electric fields (see Table 2).

We solved Eqs. (4) numerically and calculated the *I-V* dependencies for the first flux-flow branch. We used $s_c$ and $s_{ab}/g^4$ obtained from the fit of $r_{\text{ff}}(B)$, assumed $l_{ab}$=200 nm, and adjusted $g$ to obtain the *I-V* dependencies most close to experimental ones. At high fields we found the fit can be significantly improved by taking into account frequency dependence of $s_{ab}$ via Eq. (6). Obtained parameters are summarized in Table 3. Typically, far away from the boundaries and the center, the solution has the form of a regular lattice, $f_{n+1} - f_n = k$, where *k* slowly decreases with $w_E$ starting from $k = p$ at $w_E = 0$. Two lattice solutions, selected by the top and bottom boundaries, collide at the center forming defect region (shock). The first flux-flow branch terminates when $k(w_E)$ intersects the instability boundary in the $k - w_E$ plane. This corresponds to the maximum in the $j(E_z)$ dependence and happens when $w_E$ is somewhat smaller than the minimum frequency of the electromagnetic wave $w_{min}(k)$ at fixed in-plane wave vector $k = k_H$. In reduced units $w_{min}(k) = k/2$. For our parameters this corresponds to the voltage $V_{min}(H)/H \approx 0.53$ mV/junction/T. Structure of the lattice at the instability point is shown in the insert to Fig. 3.

Again, as in previous section, we show for contrast a result of the *I-V* calculation with zero in-plane dissipation, $s_{ab} = 0$ (the dashed curve in Fig 3). One can see a high disagreement with experiment in that case. Theoretical curve is running far below the experimental and the resonance upturn near $V_{max}$ is much sharper than in the experiment.

## 4. Shapiro step response of Josephson flux-flow to external microwaves

As it was predicted theoretically [8] the coherent sliding of dense lattice of Josephson lattice in layered high-$T_c$ materials should be accompanied by coherent microwave radiation with a frequency proportional to the lattice velocity. Alternatively, one can expect to find coherent DC response of Shapiro step type on the *I-V* characteristics at the Josephson flux-flow (JFF) regime under microwave radiation. Despite many studies of JFF regime in layered high-$T_c$ materials [5-7], until recently there were practically no experiments on detection of coherent emission or Shapiro steps corresponding to sliding of Josephson vortex lattice. Here we report on the observation of Shapiro-step-type response on flux-flow branch of the *I-V* characteristics of Bi-2212 "long" stacks under microwaves.

The *I-V* characteristics of samples #6 and #8 used for microwave experiments were typical to the Bi-2212 stacks with the multibranched structure [1] and critical current density $J_c = 1 - 2$ kA/cm$^2$ at 4.2 K. At fields $B > 0.03$ *T* we clearly observed flux-flow branch on the *I-V* characteristic. The properties of the flux-flow branch were similar to described in previous sections. We found that microwave irradiation induces Shapiro step structure on the flux-flow branch. That is well resolved as a series of peaks on the derivative picture *dI/dV(V)* (Fig.4). With an increase of microwave power *P* the voltage position of Shapiro steps, $V^i_{st}$, does not depend on *P* and is only a function of microwave frequency *f* in accordance with the modified Josephson relation: $V^i_{st} = i N h f / 2e$, with $i = p/q$; *p*, *q* – integers, *N* the number of the synchronized elementary junctions in the stack (Fig. 5). For one of two samples studied, #6, we found that *N* ($N = 57$) exactly corresponds to the whole number of the junctions in the stack ($N = 60\pm3$). As shown in Fig.4 the amplitudes of both, Shapiro steps $DI_c$ and critical current $I_c$ are oscillatory functions of microwave power, the oscillations of the 2$^{nd}$ step and $I_c$ peaks being in-phase, whereas oscillations of the 1$^{st}$ step and $I_c$ being in anti-phase. That resembles behaviour of Shapiro steps and critical current for conventional tunnel Josephson junctions [22]. We found also that Shapiro steps are observed at fields *B* above 1 T corresponding to the conditions of dense lattice. However, the step position at fixed microwave frequency does not depend on *B* ($B = 1$-7 T) or temperature up to 75 K.

The presence of Shapiro steps on flux-flow branch of the *I-V* characteristics proves the existence of the coherent Josephson emission induced by sliding Josephson vortex lattice. The Josephson-type relation between the voltage position of the step and external frequency points out to the in-plane coherency of moving Josepson vortex lattice, while the big coefficient *N* in Josephson relation proves the vertical coherence of the sliding Josephson vortex lattice. Our results demonstrate the principal possibility of synchronization of elementary junctions by magnetic field.

Because of the vertical synchronization of many junction one can expect to get quite high power of the microwave emission. Our rough estimation of the emitted microwave power $W$ at $f =120$ GHz as $W \sim DI_m V_{st}$ with $DI_m$ the maximum step height gives the value $\sim 10^{-6}$ Watts.

## 5. Josephson flux-flow regime in short stacks

To investigate the JFF regime in a short length limit we fabricated a set of stacks of various lengths ranged from 30 down to 1.5 microns keeping the width of all the stacks the same, close to 2 microns. We found that flux-flow regime can be achieved for all stacks including the smallest one but at different scale of magnetic field. The general feature found for the stacks of all length was the existence of threshold magnetic field $B_t$ for the flux-flow regime. The flux-flow branch appears on the *I-V* characteristic only at fields exceeding $B_t$. $B_t$ was found to increase with sample inverse length. For instance, for sample #4 with $L = 28$ μm $B_t$ was 0.03 T while for the stack #7 with $L = 3$ μm $B_t = 0.3$ T. The log-log scale dependence of threshold field on the stack inverse length is shown in Fig. 6a. That is very close to the linear dependence of $B_t(L^{-1})$ and practically does not depend on temperature at least at the interval 4.2 - 40 K. As it follows from Fig. 6a the found threshold field is close to the characteristic field $B_0$, $B_0 = F_0 / L s$, corresponding to the condition of formation of the dense JVL (dashed line in Fig 6a). The threshold field $B_t$ is found to scale with $B_0$ as $B_t = 0.63 B_0$. The flux-flow branch for short samples was rather linear with slight upturn. The maximum voltage of flux-flow branch was generally less than for long stacks but asymptotically approached that for higher fields (Fig. 6b).

For stacks shorter than 5 μm we often observed sharp kinks on the flux-flow branch of the *I-V* characteristics (Fig. 7). The voltage position of the kink, $V_k$, is field independent and for #5 is equal to ≈15 mV. Sometimes we observed also second harmonic of the kink at double voltage $2V_k$. As shown in Fig. 7 the current amplitude of the kink $DI_{k1}$ is oscillatory function of parallel magnetic field. The oscillations of $DI_{k1}(B)$ are in anti-phase with the Fraunhofer oscillations of critical current (Fig. 8), while $DI_{k2}(B)$ oscillates in phase with $I_c(B)$.

To discuss the threshold behavior of flux-flow we analyzed the value and size dependence of the first critical field $B_{c1}$ for fluxon appearance in the stack. As it was pointed out in Ref. [23] $B_{c1}$ increases with a decrease of sample size $L$ as: $B_{c1} \gg (F_0 / L^2) (l_c / l_{ab})$. We plotted that dependence in Fig.6a using the following parameters: $g = l_c / l_{ab} = 500$, $l_{ab} = 170$ nm, $s = 1.5$ nm . The dependence $B_{c1}(L^{-1})$ is shown to lie below experimental points for $B_t(L^{-1})$. That means that at fields $B_{c1} < B < B_t$ the fluxons exist in the stack but do not contribute to the flux-flow. We can consider that at those fields corresponding to dilute vortex lattice the pinning force is higher than the driving force up to the critical Josephson current across the layers. With field increase the lattice achieves dense limit and becomes rigid enough for collective motion. The increase of transverse rigidity leads to the effective reduction of pinning force, since the pinning force acting on individual fluxon will be effectively redistributed over the whole lattice. Thus we consider $B_t$ as a threshold field when the driving force exceeds the pinning force acting on JVL. That turnes out to be very near to the condition of formation of dense lattice and corresponds to the transverse fluxon density 0.63 fluxon per junction.

We can identify the kinks on the *I-V* characteristics of short stacks in Josephson flux-flow regime as the Fiske steps [24], which are known to appear due to the resonance of Josephson radiation with cavity modes in a trasmission line formed by the junction. The characteristic features of the Fiske steps are as follows:
1) The step position follows the condition $V_m = m N \Phi_0 c_0 / 2L$ with m the integer, $N$ the number of synchronized elementary junctions in the stack, $c_0$ the Swihart velocity, $L$ the stack length.
2) Voltage position of the step is independent of $B$.
3) Current amplitude of the steps, $\Delta I_m$, is periodic function on $B$ with the period $\Delta B = \Phi_0 / L s$, even steps oscillating in-phase with $I_c$, while odd step oscillations being in anti-phase with $I_c(B)$ [25].

The conditions (2-3) are valid for our kink structure (see Figs. 7,8). The estimation of voltage position also gives reasonable value $V_{k1}$ = 14.2 mV. For estimation we used the following parameters $L$ = 2 μm, $N$ = 80, $\lambda_{ab}$ = 200 nm, $s$ = 1.5 nm, $c_0/c$ =1.19 $10^{-3}$. We estimated Swihart velocity as the lowest mode valid for triangular lattice as [8] $c_0 = s c / 2 \lambda_{ab} (\varepsilon_c)^{1/2}$ with $\varepsilon_c$ = 10. Note that the resonance frequency, $c_0 / 2L$, corresponding to the 1$^{st}$ Fiske step is quite high in our case ≈ 100 GHz. We observed Fiske steps on even shorter stacks with $L$ =1.4 μm.

The first indication of Fiske steps in Bi-2212 stacked junctions has been found in [26, 27] on rather long junctions with $L$ = 20-50 μm. The steps found [27] were unstable and not well reproducible. We consider that the better conditions for Fiske step observation at short stacks are related with an increase of the resonance quality factor [25] with decreasing $L$.

## 6. Conclusions

We carried out the detailed measurements of current voltage characteristics in long Bi-2212 stacks in Josephson flux-flow regime and analyzed the data in the frame of recent theory taking into account both in-plane and the out of plane channels for dissipation in Josephson flux-flow regime. We found that the experiment is well described by that theory. From the fit we found a number of Bi-2212 parameters at T=4.2 K as $\sigma_c$, $\sigma_{ab}$ and $B_s$ as well as the temperature dependence of $\sigma_c$, which agree well with those found by other methods. From that we can conclude that in-plane dissipation plays an important role in Josephson flux-flow regime at low temperatures.

By the experiments with external microwave radiation (45 –142 GHz) we found a coherent Shapiro-step response on the *I-V* characteristic in Josephson flux-flow regime. That gives a strong evidence of the Josephson emission induced by coherently moving JVL.

For short stacks with length 1.5 – 2 microns we found the Fiske step response which appears due to the resonance of Josephson radiation with cavity modes in transmission line formed by the stack. As in the case of Shapiro step response the Fiske step position corresponds to the synchronization of all the 80 junctions in the stack.

We also found that for stacks of different length the flux-flow branch appears above some threshold field $B_t$ = 0.63 $\Phi_0 / L s$.

## Acknowledgement

We are thankful to A. M. Nikitina for providing us with Bi-2212 single crystal whiskers and S.- J. Kim for technical assistance. This work was supported by Japan

Science and Technology Corp. and by the Russian State Program on HTS (grant No. 99016).

**Table 1.** Parameters of Bi-2212 stacked junctions.

| No  | $L_a$ (μm) | $L_b$ (μm) | N  | $I_c$ (μA) | Notes |
|-----|------|------|----|------|-------|
| 4   | 28   | 2    | 90 | 550  |       |
| 2   | 7    | 2    | 84 | 300  |       |
| 7   | 3    | 2    | 81 | 140  |       |
| 5   | 2    | 2    | 80 | 18   |       |
| H-1 | 1.5  | 1.5  | 90 | 0.6  | Contains a hole D=0.2μm |
| H-3 | 1.5  | 1.5  | 80 | 0.3  | Contains a hole D=0.2μm |
| 6   | 30   | 2    | 57 | 1200 |       |
| 8   | 28   | 2.5  | 75 | 1500 |       |

**Table 2.** Meanings, definitions and practical formulas for the reduced parameters used throughout the paper. In practical formulas $f = \omega_p/2\pi$ means plasma frequency, $\rho_c$ and $\rho_{ab}$ are the components of the quasiparticle resistivity

| Notation | Meaning | Definition (CGS) | Practical formula (BSCCO) |
|----------|---------|------------------|---------------------------|
| $\omega_E$ | reduced Josephson frequency | $\dfrac{2\pi c s E_z}{\Phi_0 \omega_p}$ | $\dfrac{U[mV/junction]}{2\cdot 10^{-3} f_p[GHz]}$ |
| $k_H$ | magnetic wave vector | $\dfrac{2\pi H \gamma s^2}{\Phi_0}$ | |
| $\nu_c$ | c-axis dissipation parameter | $\dfrac{4\pi \sigma_c}{\varepsilon_c \omega_p}$ | $\dfrac{1.8\cdot 10^3}{\varepsilon_c \rho_c[\Omega\cdot cm]f_p[GHz]}$ |
| $\nu_{ab}$ | in-plane dissipation parameter | $\dfrac{4\pi \sigma_{ab} \lambda_{ab}^2 \omega_p}{c^2}$ | $\dfrac{0.79(\lambda_{ab}[\mu m])^2 f_p[GHz]}{\rho_{ab}[m\Omega\cdot cm]}$ |
| $l$ | reduced London penetration depth | $\lambda_{ab}/s$ | |

**Table 3.** Parameters of BSSCO used in calculation of the *I-V* dependencies

| Penetration depth $\lambda_{ab}$ | Anisotropy $\gamma$ | Josephson current $j_J$ | In-plane conductivity $\sigma_{ab}$ | c-axis conductivity $\sigma_c$ | Relaxation rate $1/\tau$ |
|---|---|---|---|---|---|
| 200nm | 500 | $1.7 kA/cm^2$ | $39[m\Omega\cdot cm]^{-1}$ | $2.1[k\Omega\cdot cm]^{-1}$ | $2\pi\cdot 0.55 THz$ |

**Figure captions.**

Fig. 1. Schematic view of the long Bi-2212 stack in experimental setup (a) and the *I-V* characteristics of junction #4 in magnetic field *B//b* (b).

Fig. 2. Magnetic field dependence of of the Josephson flux-flow resistance $R_{Jff}$ for long Bi-2212 stack #4 at 4.2, 20, 40, 60 and 75 K. The dashed lines are fits to Eq. (2) for each temperature. The dotted curve is a calculated dependence $R_{Jff}(B)$ at 4.2 K in a limit of zero in-plane dissipation ($\mathbf{s}_{ab} = 0$). The inset shows temperature dependence of $\mathbf{s}_c$: crosses obtained from the fit of $R_{Jff}(B)$ to Eq. 2, the line corresponds to data obtained from measurements on small mesas in zero magnetic field [20].

Fig. 3. Comparison of calculated and measured the *I-V* characteristic at *B* = 2 T. For calculation parameters see text. The inset shows vortex lattice structure near the *V* = $V_{max}$.

Fig. 4. The differential conductance of the stack #6 in flux-flow regime (*B*=2.4 T, *B//b*) as a function of microwave power of frequency 90.4 GHz. *T*=7.4 K. Conductance scale corresponds to the lowest at 8.5 dBm, the other spectra are offset vertically.

Fig. 5. Frequency dependence of the first Shapiro step position, $V^1_{st}$, as a function of the frequency of microwave field for Bi-2212 stacked junction #6. *T* = 7.4 K, *B* = 2.4 T, *B // b*. The straight line corresponds to the Josephson relation for *N* = 57.

Fig. 6. (a) Experimental dependence of threshold magnetic field for appearance of the flux-flow branch on the *I-V* characteristics as a function of the inverse length of the stack. The dased line corresponds to condition $B = \mathbf{F}_0 / L\, s$. Solid line corresponds to field $B_{c1}$: $B_{c1} = (\mathbf{F}_0\, L^2)(\mathbf{l}_c / \mathbf{l}_{ab})$. (b) Magnetic field dependence of maximum Josephson flux-flow voltage for stacks of different length.

Fig. 7. A set of the *I-V* characteristics of a short Bi-2212 stack #5 with variation of parallel magnetic field $B_{//}$ within 3.7-5.7 T. T=4.2 K. Note a Fiske step at V ≈ 15 mV.

Fig. 8. Magnetic field dependence of normalized critical current $I_c / I_{c0}$ (●) and amplitude of the 1st Fiske step $\mathbf{D}I/I_{c0}$ (o). *H // b*, T = 4.2 K. Solid line is a fit to Fraunhofer dependence $I_c / I_{c0} = \sin x / x$ with $x = \mathbf{p}\, B\, L\, s / \mathbf{F}_0$, *L* = 2 μm.

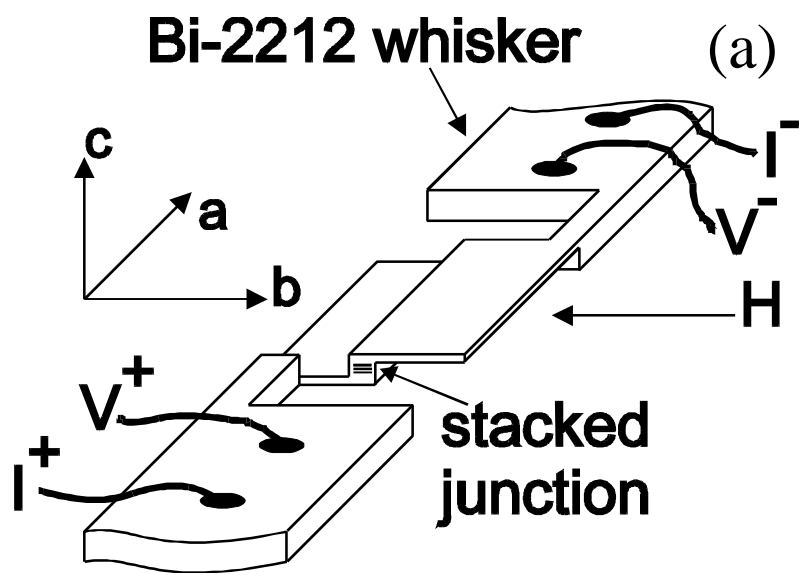
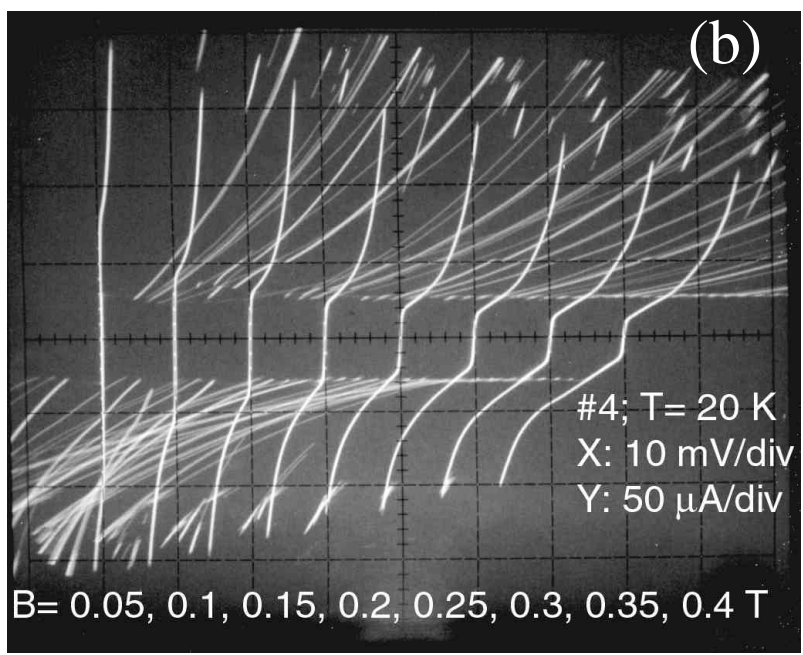

Fig. 1.

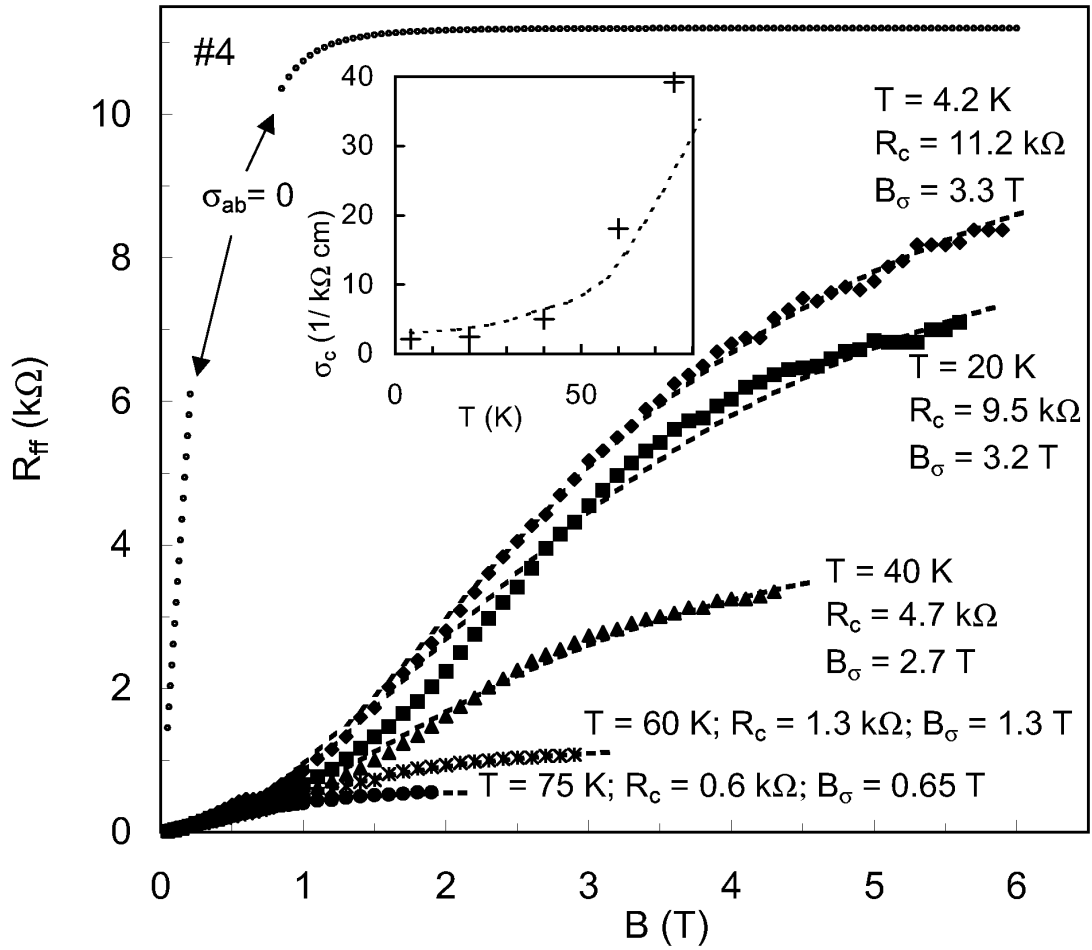

Fig. 2.

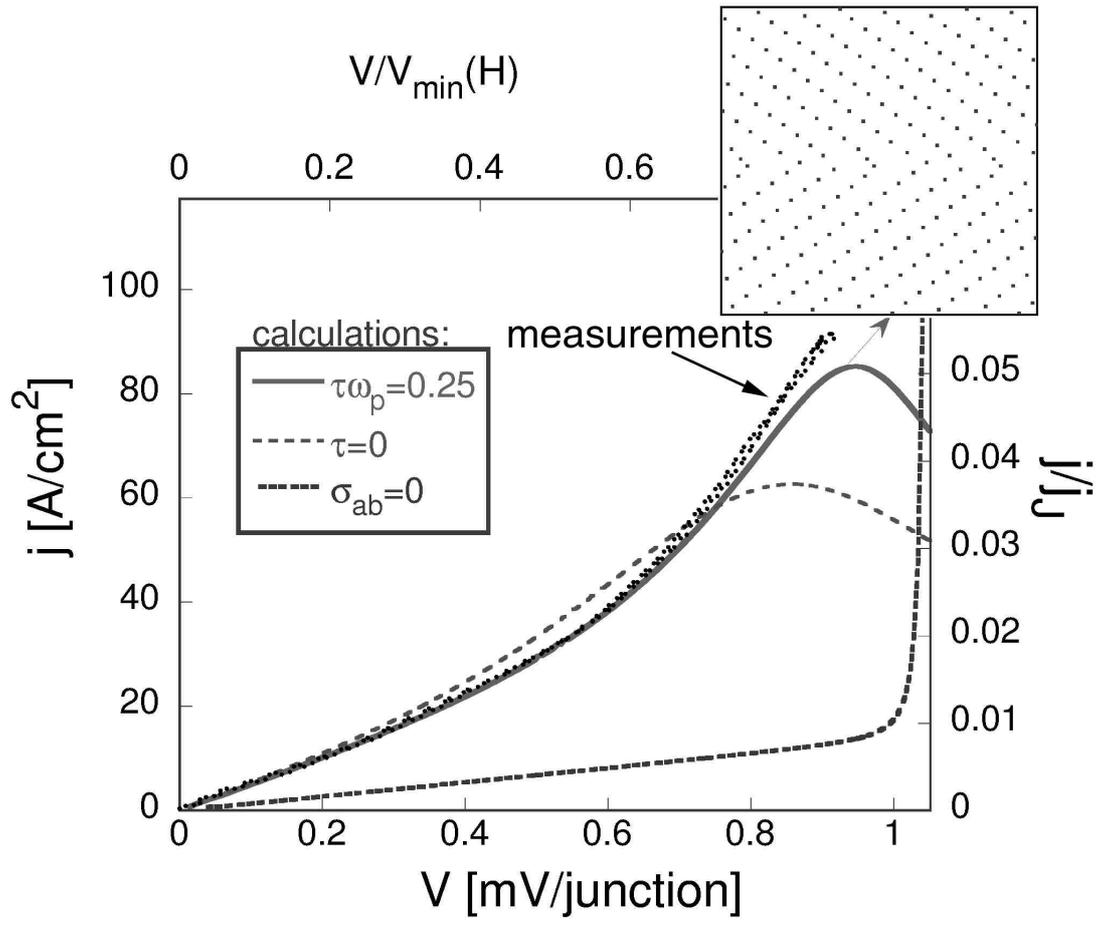

Fig. 3.

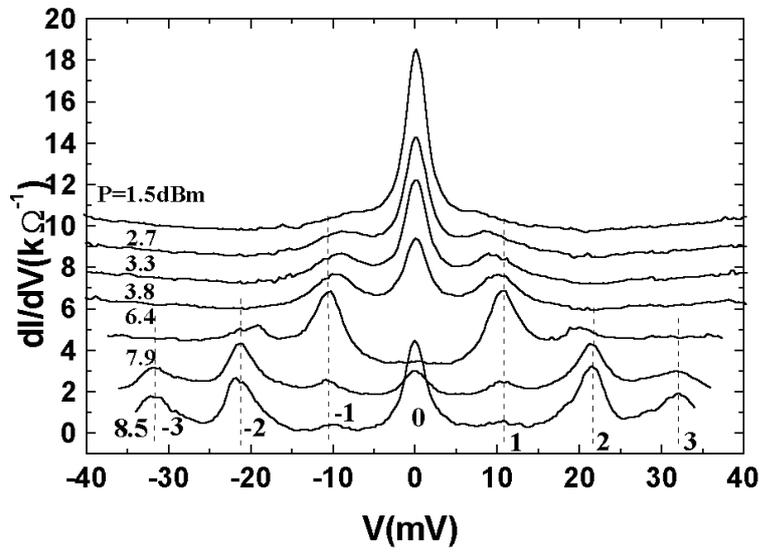

Fig. 4.

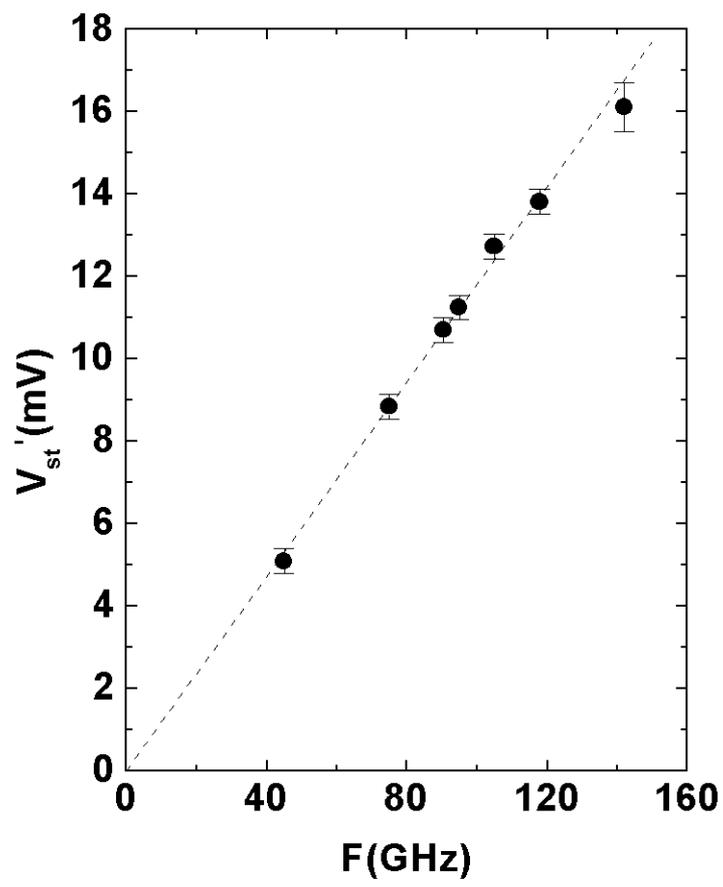

Fig. 5

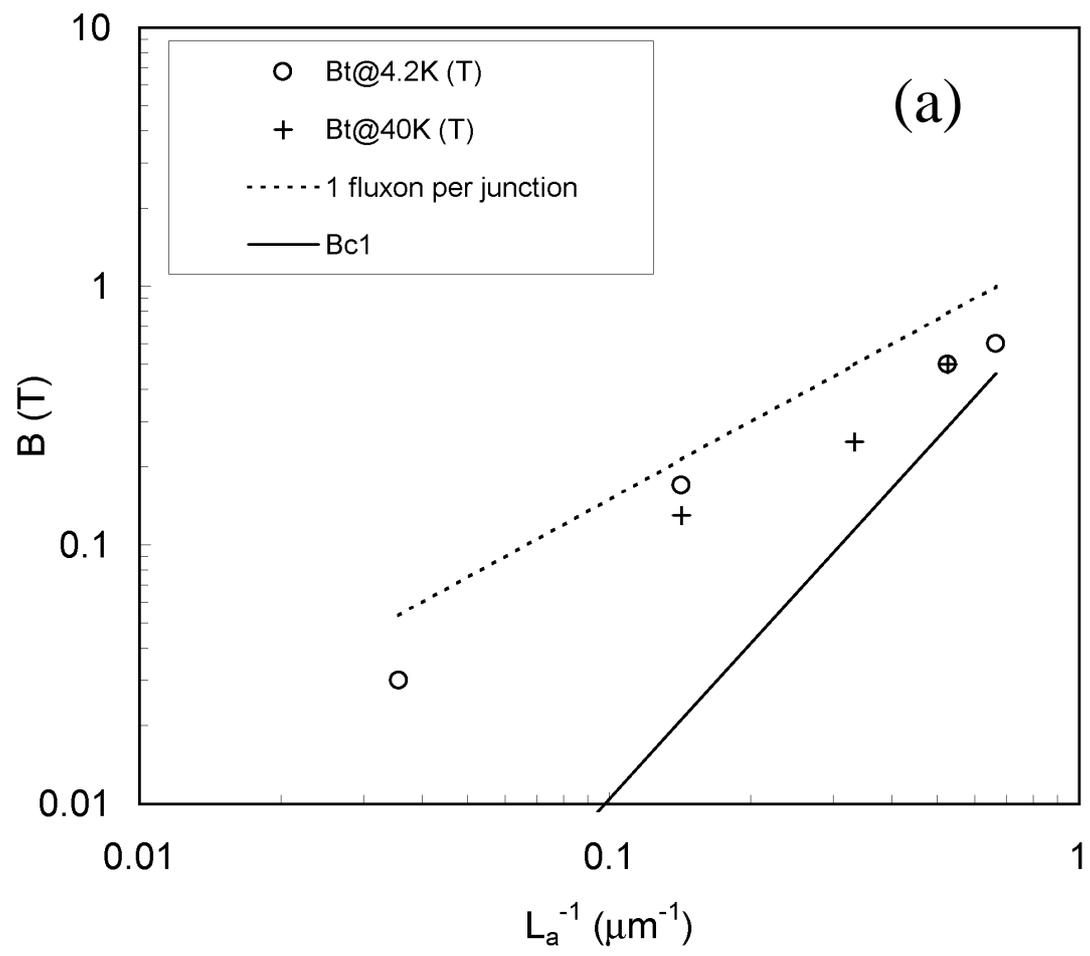

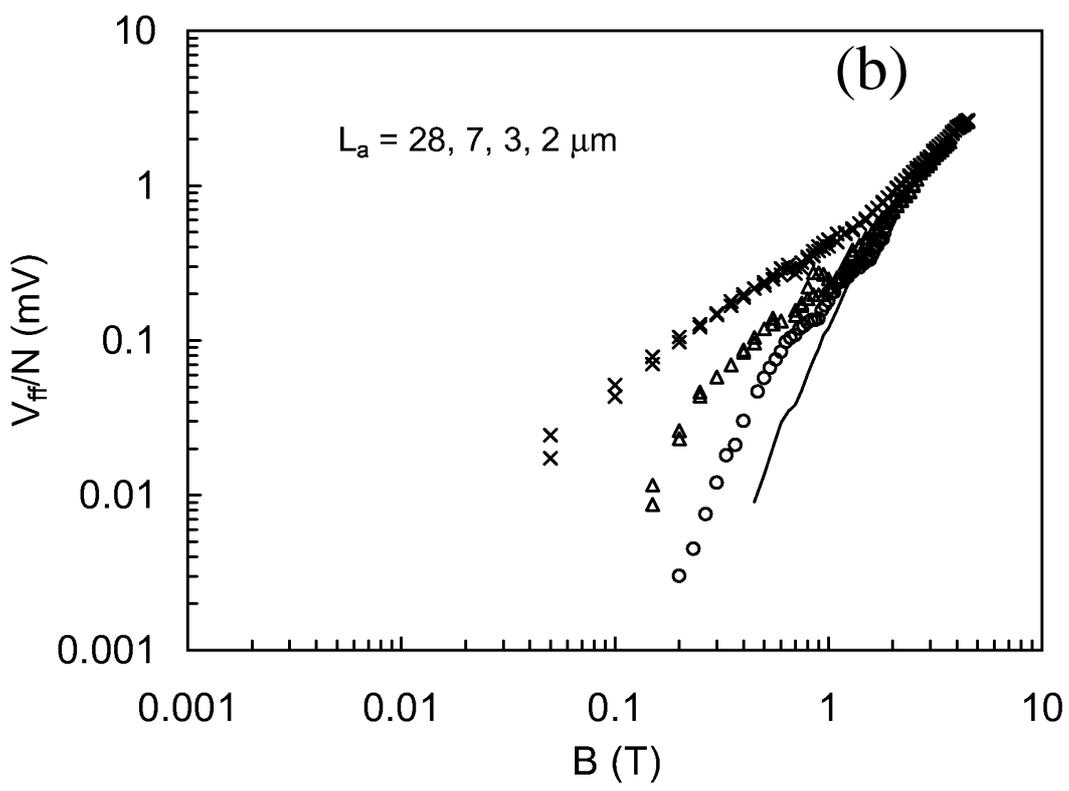

Fig. 6.

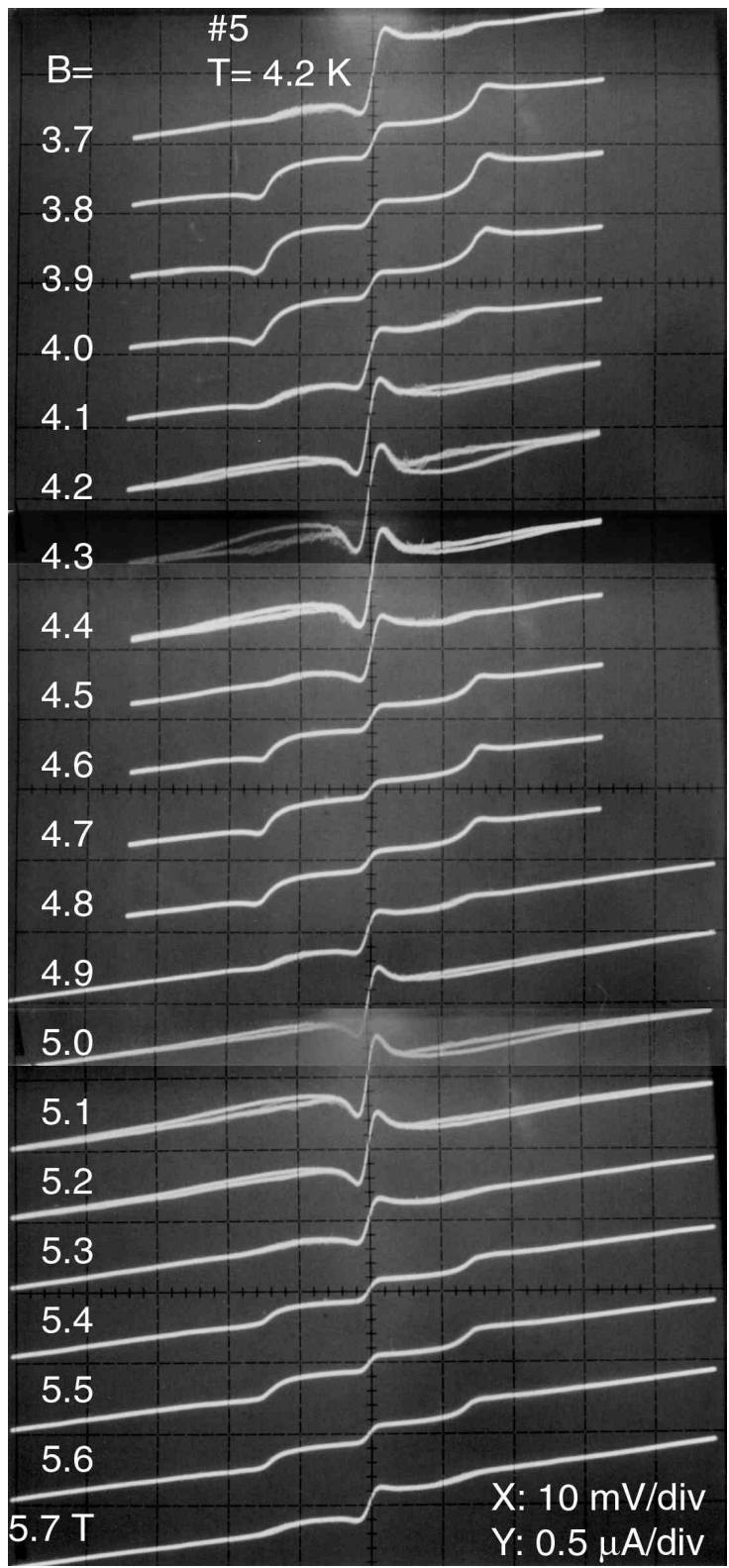

Fig. 7.

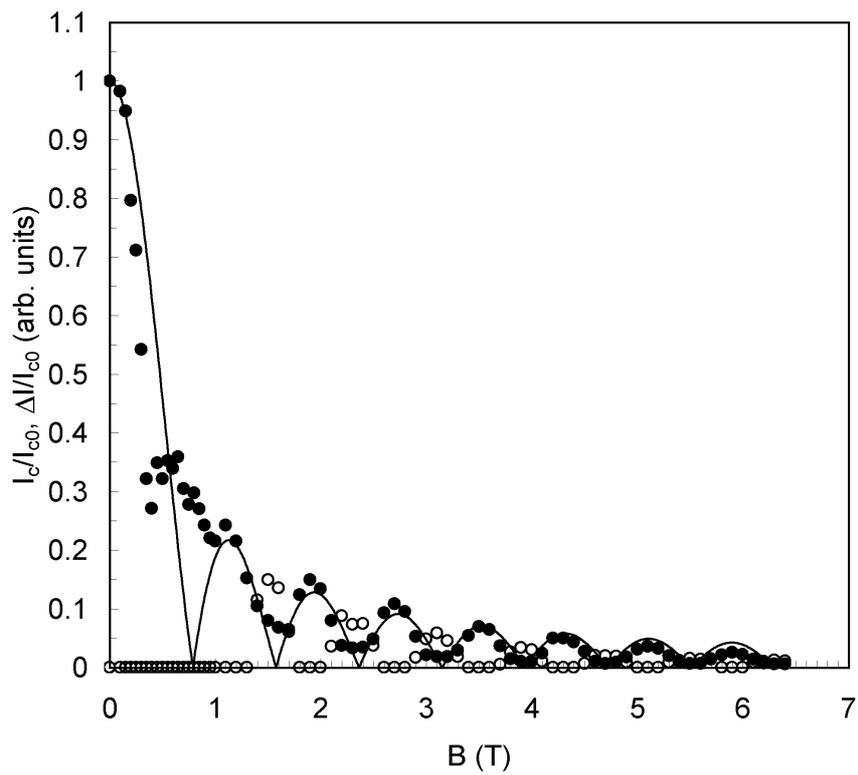

Fig. 8.